\documentclass{article}
\usepackage{amssymb}
\usepackage{amsmath}

\begin{document}

\begin{center}
{\Large Some application of difference equations in Cryptography and Coding
Theory}
\end{center}

\begin{equation*}
\end{equation*}

\begin{center}
Cristina FLAUT

{\small Faculty of Mathematics and Computer Science, Ovidius University,}

{\small Bd. Mamaia 124, 900527, CONSTANTA, ROMANIA}

{\small http://www.univ-ovidius.ro/math/}

{\small e-mail: cflaut@univ-ovidius.ro; cristina\_flaut@yahoo.com}%
\begin{equation*}
\end{equation*}
\end{center}

\textbf{Abstract.} {\small In this paper, we present some applications of a
difference equation of degree }$k${\small \ in Cryptography and Coding
Theory.}$\medskip \ $



\begin{center}
\begin{equation*}
\end{equation*}
\end{center}

\bigskip \textbf{1. Introduction}%
\begin{equation*}
\end{equation*}%
\qquad

There are many papers devoted to the study of the properties and the
applications of some particular integer sequences, as for example: Fibonacci
sequences, $p-$Fibonacci sequences, Tribonacci sequences, etc. (see [1],
[2], [4], [5], [8], [9], [10], [11]). In this paper, we generalize these
results, by considering the general case of a difference equation of degree $%
k$, we associate a matrix to such an equation and, using some properties of
these matrices, we give some applications of them in Cryptography and Coding
Theory. We generalize the notion of complete positive integers sequence,
given in [6], and a result given in [13], result which states the
representation of a natural number as a sum of nonconsecutive Fibonacci
numbers. With these results, we give an algorithm for \ messages encryption
and decryption and, in Section 3, we give an application in Coding Theory.
In Appendix, we present some MAPLE procedures. These procedures are very
helpful in the encrypting and decrypting processes.

\begin{equation*}
\end{equation*}

\begin{equation*}
\end{equation*}

\textbf{2. Some properties of a difference equation of degree }$k,k\geq 2$%
\begin{equation*}
\end{equation*}

Let  $n$ be an arbitrary positive integer and $a,b$ be arbitrary integers, $%
b\neq 0$.$~$We consider the following difference equation of degree two%
\begin{equation}
d_{n}=ad_{n-1}+bd_{n-2},d_{0}=0,d_{1}=1  \tag{2.1.}
\end{equation}%
and the attached matrix 
\begin{equation*}
D_{2}=\left( 
\begin{array}{cc}
a & b \\ 
1 & 0%
\end{array}%
\right) =\left( 
\begin{array}{cc}
d_{2} & bd_{1} \\ 
d_{1} & bd_{0}%
\end{array}%
\right) .
\end{equation*}%
It results \newline
\begin{equation*}
D_{2}^{2}=\left( 
\begin{array}{cc}
b+a^{2} & ab \\ 
a & b%
\end{array}%
\right) =\left( 
\begin{array}{cc}
d_{3} & bd_{2} \\ 
d_{2} & bd_{1}%
\end{array}%
\right) .
\end{equation*}%
Therefore, we obtain that \newline
\begin{equation}
D_{2}^{n}=\left( 
\begin{array}{cc}
d_{n+1} & bd_{n} \\ 
d_{n} & bd_{n-1}%
\end{array}%
\right) .  \tag{2.2.}
\end{equation}

$\bigskip $Since \textit{det}$D_{2}=-b$ and \textit{det}$%
D_{2}^{n}=bd_{n-1}d_{n+1}-bd_{n}^{2}=\left( -b\right) ^{n},$ the following
relation is true: 
\begin{equation}
d_{n-1}d_{n+1}-d_{n}^{2}=\left( -b\right) ^{n-1}.  \tag{2.3.}
\end{equation}%
\qquad\ \ \ \ \ \ \ \ \ \ \ \ \ \ \ \ \ \ \ \qquad\ \qquad \qquad \qquad
\qquad\ \ 

The inverse of the matrix $D_{2}^{n}$ is

\begin{equation}
D_{2}^{-n}==\frac{1}{\left( -1\right) ^{n+1}b^{n}}\left( 
\begin{array}{cc}
-bd_{n-1} & bd_{n} \\ 
d_{n} & -d_{n+1}%
\end{array}%
\right) .  \tag{2.4.}
\end{equation}

If we consider the following recurrence relation of degree three 
\begin{equation}
d_{n}=ad_{n-1}+bd_{n-2}+cd_{n-2},d_{-1}=d_{0}=d_{1}=0,d_{2}=1,c\neq 0, 
\tag{2.5.}
\end{equation}%
we have the attached matrix 
\begin{equation*}
D_{3}=\left( 
\begin{array}{ccc}
d_{3} & bd_{2}+cd_{1} & cd_{2} \\ 
d_{2} & bd_{1}+cd_{0} & cd_{1} \\ 
d_{1} & bd_{0}+cd_{-1} & cd_{0}%
\end{array}%
\right) =\left( 
\begin{array}{ccc}
a & b & c \\ 
1 & 0 & 0 \\ 
0 & 1 & 0%
\end{array}%
\right) ,
\end{equation*}%
since we can take $bd_{0}+cd_{-1}=d_{2}-ad_{1}=1.$ Therefore, using relation 
$d_{4}=a^{2}+b,$ we get

\begin{equation*}
D_{3}^{2}=\allowbreak \left( 
\begin{array}{ccc}
b+a^{2} & c+ab & ac \\ 
a & b & c \\ 
1 & 0 & 0%
\end{array}%
\right) =\allowbreak \left( 
\begin{array}{ccc}
d_{4} & bd_{3}+cd_{2} & cd_{3} \\ 
d_{3} & bd_{2}+cd_{1} & cd_{2} \\ 
d_{2} & bd_{1}+cd_{0} & cd_{1}%
\end{array}%
\right) .
\end{equation*}
From here, it results

\begin{equation}
D_{3}^{n}=\left( 
\begin{array}{ccc}
d_{n+2} & bd_{n+1}+cd_{n} & cd_{n+1} \\ 
d_{n+1} & bd_{n}+cd_{n-1} & cd_{n} \\ 
d_{n} & bd_{n-1}+cd_{n-2} & cd_{n-1}%
\end{array}%
\right) .  \tag{2.6.}
\end{equation}

Since \textit{det}$D_{3}=c$ and \textit{det}$D_{3}^{n}=$\newline
$=c^{2}\left[ d_{n}\left( d_{n}^{2}-d_{n-1}d_{n+1}\right) +d_{n-2}\left(
d_{n+1}^{2}-d_{n}d_{n+2}\right) +d_{n-1}\left( \allowbreak
d_{n-1}d_{n+2}-d_{n}d_{n+1}\right) \right] ,$ using the fact that \textit{det%
}$D_{3}^{n}=c^{n},n\geq 2$, we obtain the following relation 
\begin{equation}
d_{n}\left( d_{n}^{2}\text{-}d_{n-1}d_{n+1}\right) \text{+}d_{n-2}\left(
d_{n+1}^{2}\text{-}d_{n}d_{n+2}\right) \text{+}d_{n-1}\left( \allowbreak
d_{n-1}d_{n+2}\text{-}d_{n}d_{n+1}\right) \text{=}c^{n-2}.  \tag{2.7.}
\end{equation}

\bigskip The inverse of the matrix $D_{3}^{n}$ is 
\begin{equation}
D_{3}^{-n}=\frac{1}{c^{n-2}}\left( 
\begin{array}{ccc}
g_{11} & g_{12} & g_{13} \\ 
g_{21} & g_{22} & g_{23} \\ 
g_{31} & g_{32} & g_{33}%
\end{array}%
\right) ,  \tag{2.8.}
\end{equation}%
where \newline
$g_{11}=$ $c^{2}(-d_{n}d_{n-2}+d_{n-1}^{2})\allowbreak $, $%
g_{21}=-c(-d_{n}^{2}+d_{n-1}d_{n+1}),$\newline
$g_{31}=-(bd_{n}^{2}+cd_{n}d_{n-1}-bd_{n-1}d_{n+1}-cd_{n+1}d_{n-2}),$\newline
$%
g_{12}=-c^{2}(d_{n}d_{n-1}-d_{n+1}d_{n-2}),g_{22}=c(-d_{n}d_{n+1}+d_{n-1}d_{n+2}),
$\newline
$g_{32}=cd_{n}^{2}+bd_{n}d_{n+1}-bd_{n-1}d_{n+2}-cd_{n-2}d_{n+2},$\newline
$g_{13}=c^{2}(d_{n}^{2}-d_{n-1}d_{n+1}),g_{23}=c(-d_{n}d_{n+2}+d_{n+1}^{2}),$%
\newline
$g_{33}=bd_{n}d_{n+2}-cd_{n}d_{n+1}-bd_{n+1}^{2}+cd_{n-1}d_{n+2}.$

$\allowbreak $

Now, we consider the general $k-$terms recurrence, $n,k\in \mathbb{N},k\geq
2,n\geq k,$%
\begin{equation}
d_{n}\text{=}a_{1}d_{n-1}\text{+}a_{2}d_{n-2}\text{+}...\text{+}%
a_{k}d_{n-k},d_{0}\text{=}d_{1}\text{=...=}d_{k-2}\text{=}0,d_{k-1}\text{=}%
1,a_{k}\neq 0  \tag{2.9.}
\end{equation}%
and the matrix $D_{k}\in \mathcal{M}_{k}\left( \mathbb{R}\right) ,$ 
\begin{equation}
D_{k}=\left( 
\begin{array}{ccccc}
a_{1} & a_{2} & a_{3} & ... & a_{k} \\ 
1 & 0 & 0 & ... & 0 \\ 
0 & 1 & 0 & ... & 0 \\ 
... & ... & ... & ... & ... \\ 
0 & 0 & ... & 1 & 0%
\end{array}%
\right) ,  \tag{2.10.}
\end{equation}%
(see [7]).\medskip 

\textbf{Proposition 2.1.} \textit{With the above notations, the following
relations are true:} \newline
1) 
\begin{equation}
D_{k}=\ \left( 
\begin{array}{ccccc}
d_{k} & \overset{k-1}{\underset{i=1}{\sum }}a_{i+1}d_{k-i} & \overset{k-2}{%
\underset{i=1}{\sum }}a_{i+2}d_{k-i} & ... & a_{k}d_{k-1} \\ 
d_{k-1} & \overset{k-1}{\underset{i=1}{\sum }}a_{i+1}d_{k-i-1} & \overset{k-2%
}{\underset{i=1}{\sum }}a_{i+2}d_{k-i-1} & ... & a_{k}d_{k-2} \\ 
d_{k-2} & \overset{k-1}{\underset{i=1}{\sum }}a_{i+1}d_{k-i-2} & \overset{k-2%
}{\underset{i=1}{\sum }}a_{i+2}d_{k-i-2} & ... & a_{k}d_{k-3} \\ 
... & ... & ... & ... & ... \\ 
d_{1} & \overset{k-1}{\underset{i=1}{\sum }}a_{i+1}d_{k-i-k+1} & \overset{k-2%
}{\underset{i=1}{\sum }}a_{i+2}d_{-i+1} & ... & a_{k}d_{0}%
\end{array}%
\right) .  \tag{2.11.}
\end{equation}

\text{2) For }$n\in \mathbb{Z},n\geq 1,$\text{ we have that }

\begin{equation}
D_{k}^{n}\text{=}\left( 
\begin{array}{ccccc}
d_{n+k-1} & \overset{k-1}{\underset{i=1}{\sum }}a_{i+1}d_{n+k-i-1} & \overset%
{k-2}{\underset{i=1}{\sum }}a_{i+2}d_{n+k-i-1} & ... & a_{k}d_{n+k-2} \\ 
d_{n+k-2} & \overset{k-1}{\underset{i=1}{\sum }}a_{i+1}d_{n+k-i-2} & \overset%
{k-2}{\underset{i=1}{\sum }}a_{i+2}d_{n+k-i-2} & ... & a_{k}d_{n+k-3} \\ 
d_{n+k-3} & \overset{k-1}{\underset{i=1}{\sum }}a_{i+1}d_{n+k-i-3} & \overset%
{k-2}{\underset{i=1}{\sum }}a_{i+2}d_{n+k-i-3} & ... & a_{k}d_{n+k-4} \\ 
... & ... & ... & ... & ... \\ 
d_{n} & \overset{k-1}{\underset{i=1}{\sum }}a_{i+1}d_{n+k-i-k} & \overset{k-2%
}{\underset{i=1}{\sum }}a_{i+2}d_{n-i} & ... & a_{k}d_{n-1}%
\end{array}%
\right) .  \tag{2.12.}
\end{equation}%
\newline

\textbf{Proof. }1) It is obvious, using relation $\left( 2.9\right) $.
Indeed, $d_{k}=a_{1}.$ Making computations, we obtain 
\begin{equation*}
\overset{k-1}{\underset{i=1}{\sum }}%
a_{i+1}d_{k-i}=a_{2}d_{k-1}+a_{3}d_{k-2}+...+a_{k}d_{1}=a_{2},
\end{equation*}%
\begin{equation*}
...
\end{equation*}%
\begin{equation*}
\overset{k-1}{\underset{i=1}{\sum }}a_{i+1}d_{k-i-2}\text{=}%
a_{2}d_{k-3}+a_{3}d_{k-4}+...+a_{k}d_{-1}\text{=}d_{k-1}-a_{1}d_{k-2}=1,
\end{equation*}
and so on.

2) \ We use induction. For $n=1$, the relation is true. Let $%
D_{k}^{n+1}=\left( h_{ij}\right) _{i,j\in \{1,2,...,k\}}$. Since $%
D_{k}^{n+1}=D_{k}^{n}D_{k}$, assuming that the statement is true for $n$ and
using relation $\left( 2.10\right) ,~$it results the following elements for
the matrix $D_{k}^{n+1}:$\newline
\begin{equation*}
h_{11}\text{=}a_{1}d_{n+k-1}\text{+}\overset{k-1}{\underset{i=1}{\sum }}%
a_{i+1}d_{n+k-i-1}\text{=}a_{1}d_{n+k-1}+a_{2}d_{n+k-2}+...+a_{k}d_{n}\text{=%
}d_{n+k}\newline
,
\end{equation*}%
\begin{equation*}
h_{12}\text{=}a_{2}d_{n+k-1}+\overset{k-2}{\underset{i=1}{\sum }}%
a_{i+2}d_{n+k-i-1}\text{=}a_{2}d_{n+k-1}\text{+}a_{3}d_{n+k-2}\text{+...+}%
a_{k}d_{n-1}\text{=}\overset{k-1}{\underset{i=1}{\sum }}a_{i+1}d_{n+k-i},
\end{equation*}

\begin{equation*}
...
\end{equation*}%
\begin{equation*}
h_{1k}=a_{k}d_{n+k-1},
\end{equation*}%
\begin{equation*}
h_{21}\text{=}a_{1}d_{n+k-2}\text{+}\overset{k-1}{\underset{i=1}{\sum }}%
a_{i+1}d_{n+k-i-2}\text{=}a_{1}d_{n+k-2}+a_{2}d_{n+k-3}+...+a_{k}d_{n-1}%
\text{=}d_{n+k-1},
\end{equation*}%
\begin{equation*}
h_{22}\text{=}a_{2}d_{n+k-2}\text{+}\overset{k-2}{\underset{i=1}{\sum }}%
a_{i+2}d_{n+k-i-2}\text{=}a_{2}d_{n+k-2}\text{+}a_{3}d_{n+k-3}\text{+...+}%
a_{k}d_{n}\text{=}\overset{k-1}{\underset{i=1}{\sum }}a_{i+2}d_{n+k-i-1},
\end{equation*}%
etc.$~\Box \medskip $

\textbf{Remark 2.2. ([7]) }The following statements are true:\newline
i) \textit{det}$D_{k}=\left( -1\right) ^{k+1}a_{k}$ and 
\begin{equation*}
D_{k}^{-1}=\frac{1}{a_{k}}\left( 
\begin{array}{ccccc}
0 & a_{k} & 0 & ... & 0 \\ 
1 & 0 & a_{k} & ... & 0 \\ 
... & ... & ... & ... & ... \\ 
0 & 0 & 0 & ... & 1 \\ 
1 & -a_{1} & ... & -a_{k-2} & -a_{k-1}%
\end{array}%
\right) .
\end{equation*}%
\newline
ii) $D_{k}^{n}=a_{1}D_{k}^{n-1}+a_{2}D_{k}^{n-2}+...+a_{k}D_{k}^{n-k}.%
\medskip $\newline

\textbf{Remark 2.3.} From the above, we have that

\begin{equation}
\text{\textit{det}}D_{k}^{n}=\left( -1\right) ^{\left( k+1\right)
n}a_{k}^{n}.  \tag{2.13.}
\end{equation}%
Relations 2.3, 2.7 and 2.13 are Cassini's type relations.

\textbf{%
\begin{equation*}
\end{equation*}%
}

3. \textbf{Applications in cryptography and coding theory}%
\begin{equation*}
\end{equation*}

A sequence of positive integers, $\left( d_{n}\right) ,n\geq 0,$ is \textit{%
complete} if and only if each natural number $n$ can be written under the
form $\underset{i=1}{\overset{m}{\sum }}c_{i}d_{i},$ where $c_{i}$ is either
zero or one (see [6]). In [3], the author proved that a nondecreasing
sequence of natural numbers, with $d_{1}=1,$ is complete if and only if the
following relation 
\begin{equation*}
d_{k+1}\leq 1+\underset{i=1}{\overset{k}{\sum }}d_{i},
\end{equation*}%
is true for $k\in \{1,2,3,.....\}.$

There are sequences of positive integers which are complete and sequences
which are not complete. An example of complete sequence is the sequence of
Fibonacci numbers. In [13], the author proved that each natural number can
be written as a unique sum of non-consecutive Fibonacci numbers, therefore
the sequence of Fibonacci numbers is complete. An example of not complete
sequence was done in [12]. The sequence $\left( d_{n}\right) ,n\geq 0,$
where $d_{n}=4d_{n-1}+3d_{n-2},d_{0}=0,d_{1}=1,$ is not complete.

In the following, we will generalize the notion of complete sequence and we
will give applications of this new notion.\medskip 

\textbf{Definition 3.1.} A sequence of positive integers, $\left(
d_{n}\right) ,n\geq 0,$ is called \textit{general} \textit{complete }(or 
\textit{g-complete}) if and only if for each natural number $n\geq 0,$ there
is a natural number $q,$ $q\geq 0,$ such that $n$ can be written, in a
unique way, under the form 
\begin{equation*}
n=d_{0}a_{0}+a_{1}d_{1}+...+a_{q}d_{q},a_{0},...a_{q}\in \mathbb{N}%
,a_{q}\neq 0.
\end{equation*}

\textbf{Theorem 3.2.}\textit{\ The sequence of positive integers }$\left(
d_{n}\right) ,n\geq 0,$\textit{\ generated by the difference equation} $%
\left( 2.9\right) $ \textit{is g-complete.\medskip }

\textbf{Proof.} We consider the sequence of positive integers $\left(
d_{n}\right) _{n\geq 0}$ generated by the difference equation given in $%
\left( 2.9\right) $. Let $n,q$ be the natural numbers, with $q\geq k\geq 2,~$%
such that the term $d_{q},$ of the sequence $\left( d_{n}\right) _{n\geq 0},$
satisfies the condition $d_{q}\leq n<d_{q+1}.$ We use the Quotient Remainder
Theorem. Therefore, we obtain the natural numbers $c_{q},c_{q-1},....c_{k-1}$
and $r_{q},r_{q-1},...r_{k}$ such that\newline
\begin{equation*}
n=d_{q}c_{q}+r_{q},0\leq r_{q}<d_{q},\newline
\end{equation*}%
\begin{equation*}
r_{q}=d_{q-1}c_{q-1}+r_{q-1},0\leq r_{q-1}<d_{q-1},\newline
\end{equation*}%
\begin{equation*}
r_{q-1}=d_{q-2}c_{q-2}+r_{q-2},0\leq r_{q-2}<d_{q-2},\newline
\end{equation*}%
\begin{equation*}
........\newline
\end{equation*}%
\newline
\begin{equation*}
r_{k+1}=d_{k}c_{k}+r_{k},0\leq r_{k}<d_{k},\newline
\end{equation*}%
\begin{equation*}
r_{k}=d_{k-1}c_{k-1},
\end{equation*}%
with $c_{k-1}=r_{k},$ since $d_{k-1}=1.$\newline
It results that the number $n$ can be written, in a unique way, under the
form 
\begin{equation}
n=d_{q}c_{q}+d_{q-1}c_{q-1}+d_{q-2}c_{q-2}+...+d_{k}c_{k}+d_{k-1}c_{k-1}, 
\tag{3.1.}
\end{equation}

Indeed, we can't have $k$ terms in the relation $\left( 3.1\right) ,$\newline
$d_{s},d_{s+1},...d_{s+k-1},$ with $s+k-1\leq q$ such that $%
c_{s}d_{s}+c_{s+1}d_{s+1}+...+c_{s+k-1}d_{s+k-1}=d_{s+k}.$ In relation $%
\left( 3.1\right) $, if $s+k<q,$ then the coefficient of $d_{s+k}$ is $%
c_{s+k}+1$. From here, we get 
\begin{equation*}
r_{s+k+1}=\left( c_{s+k}+1\right) d_{s+k}+r_{s+k}-d_{s+k}.
\end{equation*}%
Since $r_{s+k}-$ $d_{s+k}<0,$ we obtain a contradiction with the Quotient
Remainder Theorem.$~\Box \medskip $

The above Theorem extend the result obtained in [13] for Fibonacci numbers
to a difference equation of degree $k\,$, equation defined by $\left(
2.9\right) $. 
\begin{equation*}
\end{equation*}

\textbf{3.1. An application in Cryptography}%
\begin{equation*}
\end{equation*}

In [8], [10], [11], [12], were described applications of Fibonacci numbers, $%
k-$Fibonacci numbers or Tribonacci numbers in Cryptography. In the
following, by using Theorem 3.2 and the terms $\left( d_{n}\right) _{n\geq 0}
$ of the difference equation given by the relation $\left( 2.9\right) $, we
will present a new method for encrypting and dencrypting messages. This new
method give us a lot of possibilities for finding the encryption and
decryption keys, having the advantage that each natural number $n$ has a
unique representation by using the terms of such a sequence. It results that
a number cannot have the same encrypted value as another number, therefore
the obtained encrypted texts are hard to break.\medskip \medskip 

\textbf{The Algorithm}\medskip

Let $\mathcal{A}$ be an alphabet with $N$ letters, labeled from $0$ to $N-1,$
$m$ be a plain text and $n$ be a number obtained by using the label of the
letters from the plain text $m.$ We split the text $m$ in blocks with the
same length, $m_{1},m_{2},...,m_{r}.$ To these blocks correspond the numbers 
$n_{1},n_{2},...,n_{r}.$ The numbers $n_{i},i\in \{1,2,...,r\},$ will be
encrypted using the following procedure.

\bigskip

\textit{The encrypting\medskip }

1. We consider the natural numbers $k\geq 2,a_{1},a_{2},...,a_{k};$

2. We consider the difference equation of degree $k,$ given by the relation $%
\left( 2.9\right) ,$ with the above coefficients $a_{1},a_{2},...,a_{k}.$

3. We compute the elements $d_{k},d_{k+1},...,d_{q},~$such that $d_{q}\leq
n_{i}<d_{q+1},i\in \{1,2,...,r\},$ as in the proof of Theorem 3.2. We denote 
$s=q-k+2.$

4. We obtain the secret encryption and decryption key $\left(
a_{1},a_{2},...,a_{k},s\right) .$

5. We compute the numbers $c_{iq},c_{i(q-1)},...,c_{i(k-1)}$ such that 
\begin{equation*}
n_{i}=d_{q}c_{iq}+d_{q-1}c_{i(q-1)}+d_{q-2}c_{i(q-2)}+...+d_{k}c_{ik}+d_{k-1}c_{i(k-1)},
\end{equation*}
as in relation $\left( 3.1\right) .$

6. We obtain $\left( c_{iq},c_{i(q-1)},...,c_{i(k-1)}\right) ,$ the labels
for the encrypted text. If all $c_{ij}\leq N-1,$ then we will put in the
cipher text the letter $L\,_{j},$ from the alphabet $\mathcal{A},$ which has
the label $c_{ij}.$ If there are $c_{ij}>N-1,$ then we will put $c_{ij}~$in
the cipher text. If there are more than one $c_{ij}$ such that $c_{ij}>N-1,$
namely $\{c_{ij_{1}},...c_{ij_{t}}\},~$we will count the biggest number of
decimals of these elements. Assuming that $c_{ij_{v}}$ has the biggest
number of decimals, $0\leq v\leq t,$ we will put zeros to the left side of
the $c_{ij_{p}},p\neq v,$ such that all new obtained $c_{ij_{l}}~$have the
same number of decimals. In this way, we obtain the encrypted text $T_{i}$.

7. Using the above steps for all blocks, we obtain the cipher text, by joint
the cipher texts $T_{1}T_{2}...T_{r}$.\medskip

\textit{The decrypting\medskip }

1. We use the key $\left( a_{1},a_{2},...,a_{k},s\right) .$ The last number $%
s$ is the length of encoded blocks.

2. We split the cipher text in blocks of length $s$ and we consider their
labels $\left( c_{iq},c_{i(q-1)},...,c_{i(k-1)}\right) .~$We compute the
elements $d_{k},d_{k+1},...d_{q}$, therefore the corresponding plain text
for each block is $%
n_{i}=d_{q}c_{iq}+d_{q-1}c_{i(q-1)}+d_{q-2}c_{i(q-2)}+...+d_{k}c_{ik}+d_{k-1}c_{i(k-1)}. 
$

3. If there are numbers $\{c_{ij_{1}},...c_{ij_{t}}\}$ which are greater than%
$~N-1,$ these numbers appear with the same number of digits or with new
labels, therefore it is easy to find the real message.\medskip 

\textbf{Example 3.3.} 1) We consider an alphabet with 27 letters: A, B,
C,.....Z, x, where \textquotedblright x\textquotedblright\ represent the
blank space, labeled with $0,1,2,...,26.$ We want to encrypt the message
\textquotedblright JOHNxHASxAxDOG\textquotedblright . We split this message
in the following blocks: JOHN, xHAS, xAxD, OGxx. We added two characters at
the end of the last block to obtain a block with $4$ letters. We consider a
difference equation of degree four, therefore $k=4,$ and $%
a_{1}=18,a_{2}=10,a_{3}=13,a_{4}=3$. Therefore, the encoding key is SKND, or
18101303. If, for example \thinspace $a_{1}=182,$ the key will be
182010013003.

To the first block, JOHN, will correspond the number $n_{1}=9140713.$ We
have $d_{0}=d_{1}=d_{2}=0,d_{3}=1$. We obtain the following terms \newline
$d_{4}=18,d_{5}=334,d_{6}=6205,$ $d_{7}=115267,d_{8}=2141252.$ Since $%
d_{8}\leq 9140713\leq d_{9},$ we obtain the key$~\left( 18,10,13,3,6\right) $%
.\newline
We have\newline
$n_{1}=\mathbf{4}\cdot d_{8}+575705,$\newline
$575705=\mathbf{4}\cdot d_{7}+114637,\newline
114637=\mathbf{18}\cdot d_{6}+2947,\newline
2947=\mathbf{8}\cdot d_{5}+275,\newline
275=\mathbf{15}\cdot d_{4}+5,\newline
5=\mathbf{5}\cdot d_{3},$ therefore, we obtain the labels for the encrypted
text $\left( 4,4,18,8,15,5\right) $. It results that the first block is
encrypted in the word EESIPF. To do this faster, we can use the below MAPLE
procedures:
\begin{verbatim}
> crypt(JOHN); 
 9140713
>lineq(4, 18, 10, 13, 3, 9140713);
6
40418081505
> decrypt(40418081505);
EESIPF 
\end{verbatim}

For the second block, xHAS, we get the number $n_{2}=26070018.$ Using the
same algorithm, we obtain the following labels for the encrypted text $%
\left( 12,3,4,13,1,13\right) .$ The second block is encrypted in the word
MDENBN. To ease calculations, we can use the below MAPLE procedures:
\begin{verbatim}
> crypt(xHAS); 
 26070018
>lineq(4, 18, 10, 13, 3, 26070018);
6
120304130113
> decrypt(120304130113);
MDENBN 
\end{verbatim}

To the third block, xAxD, corresponds the number $n_{3}=26002603.$ We obtain%
\newline
the labels for the encrypted text $\left( 12,2,12,7,13,13\right) .$ It
results that the third block is encrypted in the word MCMHNN, by using the
below MAPLE procedures:
\begin{verbatim}
> crypt(xAxD); 
 26002603
>lineq(4, 18, 10, 13, 3, 26002603);
6
120212071313
> decrypt(120212071313);
MCMHNN
\end{verbatim}

To the last block, OGxx, corresponds the number $n_{4}=14062626$. We get%
\newline
the labels for the encrypted text $\left( 6,10,10,1,3,6\right) .$ The fourth
block is encrypted in the word GKKBDG, as we can see by using the below
MAPLE procedures:
\begin{verbatim}
> crypt(OGxx); 
 14062626
>lineq(4, 18, 10, 13, 3, 14062626);
6
61010010306
> decrypt(61010010306);
GKKBDG
\end{verbatim}

Therefore, the encrypted text is EESIPFMDENBNMCMHNNGKKBDG.

For decoding, we use the key $\left( 18,10,13,3,6\right) .~$We split the
message in blocks of length $6$: EXKLFC, MDENBN, MCMHNN, GKKBDG. For this,
we can use the below MAPLE procedure:
\begin{verbatim}
> split(EESIPFMDENBNMCMHNNGKKBDG, 6);
[EESIPF MDENBN MCMHNN GKKBDG] 
  
\end{verbatim}

We compute $d_{4}=18,d_{5}=334,d_{6}=6208,$ $d_{7}=115267,d_{8}=2141252$.
For EESIPF, we obtain $n_{1}=2141252\cdot 4+115267\cdot 4+6205\cdot
18+334\cdot 8+18\cdot 15+5\cdot 1=9140713,$ that means we get the word JOHN,
as we can see below:
\begin{verbatim}
> delineq(EESIPF,18, 10, 13, 3);
9140713
\end{verbatim}

2) If we consider $k=4,$ $a_{1}=18,a_{2}=4,a_{3}=13,a_{4}=3$, then, the
encoding key is SEND, or 18041303. \ In this situation, we have the
following terms $d_{4}=18,d_{5}=328,d_{6}=5989,$ $d_{7}=109351,d_{8}=1996592,
$ and the encrypted message is E28OQAPNBAPCPNAHPDKHAOIAM.

3) If the alphabet has 27 letters and we received \ X032121MBD, as an
encrypted block, and we know that this block has \textquotedblright
length\textquotedblright\ $6,$ we obtain the labels $\left(
23,32,121,12,1,3\right) ,$ that means $23032121012001003.$ Indeed, $23$ is
label for X, the next two labels are $032$ and $121,$ since we know that the
labels have the same number of digits, in our case $3,$ since the block has
level $6$. For this situation, the procedure \textit{decrypt} was modified
in \textit{decryptM} \ by inserting new characters \textit{y} for $032$ and 
\textit{z} for $121.$
\begin{verbatim}
> decryptM(23032121012001003);
XyzMBD
\end{verbatim}

If the above block has length $7,$ then the labels are $\left(
23,3,21,21,12,1,3\right) $.

\begin{equation*}
\end{equation*}

\textbf{3.2. An application in Coding Theory}%
\begin{equation*}
\end{equation*}

In [1], [4], [9], were presented some applications of Fibonacci numbers\ and
Fibonacci $p-$numbers in Coding Theory. With these elements, were defined
some special matrices used for sending messages. In the following, we
generalized these results, using the matrices given in relations $\left(
2.11\right) \,\ $and $\left( 2.12\right) $ as a coding matrices for sending
messages. These matrices are associated to difference equation of order $k,$ 
$k$ a natural number, $k\geq 2,$ given by the relation $\left( 2.9\right) ,$
with $a_{1},...a_{k}$ positive integers$.~$We attach to this equation the
following equation of degree $k$%
\begin{equation}
x^{k}-a_{1}x^{k-1}-a_{2}x^{k-2}-...-a_{k}=0  \tag{3.2.}
\end{equation}%
and we assume that equation $\left( 3.2\right) $ has $k$ distinct real roots 
$\alpha _{1},\alpha _{2},...,\alpha _{k},$ such that $\alpha _{1}>1$ and $%
\alpha _{1}>\left\vert \alpha _{i}\right\vert ,i\in \{2,...,k\}.$

We assume that the initial message is represented as a square matrix $%
M=\left( m_{ij}\right) _{i,j\in \{1,2,...,k\}}~$of order $k,$ with $%
m_{ij}\geq 0,i,j\in \{1,2,...,k\}.$ Considering that $D_{k}^{n}=\left(
h_{ij}\right) _{i,j\in \{1,2,...,k\}}$ is the coding matrix and $D_{k}^{-n}$
is the decoding matrix, we will use the relation$\ $%
\begin{equation}
D_{k}^{n}\cdot M=E,  \tag{3.3.}
\end{equation}%
as encoding transformation and the relation 
\begin{equation}
M=D_{k}^{-n}\cdot E,  \tag{3.4.}
\end{equation}%
as decoding transformation.\medskip\ We define $E$ as code-message matrix.

\textbf{Theorem 3.4.} \textit{Withe the above notations, the following
statement is true}

\begin{equation}
\frac{e_{ij}}{e_{(i+r)j}}\thickapprox \alpha _{1}^{r},r\in \{1,2,...,k-1\}, 
\tag{3.5.}
\end{equation}%
\textit{for} $i+r\leq k.\medskip $

\textbf{Proof.} Since $a_{1},\alpha _{2},...,\alpha _{k}$ are roots of the
equation $\left( 3.2\right) ,$ we have 
\begin{equation}
d_{n}=A_{1}\alpha _{1}^{n}+A_{2}\alpha _{2}^{n}+...+A_{k}\alpha _{k}^{n}, 
\tag{3.6.}
\end{equation}%
with coefficients $A_{i}\in \mathbb{R},i\in \{1,2,...,k\},$ obtained from
the initial conditions of difference equation $\left( 2.9\right) .$ Using
relation $\left( 3.3\right) ,$ we get $e_{ij}=\overset{k}{\underset{t=1}{%
\sum }}h_{it}m_{tj}$ $.$ From relation $\left( 2.12\right) ,$ it results
that the elements \thinspace $h_{ij}$ are non-zero elements and are
represented as a linear combination of the terms $%
d_{n+k-i},d_{n+k-i-1},...,d_{n-i+1},$%
\begin{equation*}
h_{ij}=\overset{k}{\underset{t=1}{\sum }}B_{tj}d_{n+k-i-t+1},
\end{equation*}%
$B_{tj}\in \{0,1,a_{1},a_{2},...a_{k}\}.$ For example, 
\begin{equation*}
h_{i1}=d_{n+k-i}=\overset{k}{\underset{t=1}{\sum }}B_{t1}d_{n+k-i-t+1},
\end{equation*}%
where $B_{11}=1,B_{t1}=0,t\in \{2,...k\},$ 
\begin{equation*}
h_{i2}=\overset{k-1}{\underset{t=1}{\sum }}a_{t+1}d_{n+k-t-i}=\overset{k}{%
\underset{t=1}{\sum }}B_{t2}d_{n+k-i-t+1},
\end{equation*}%
where $B_{12}=0,B_{22}=a_{2},B_{32}=a_{3},...B_{k2}=a_{k},~$etc.

Therefore, we obtain 
\begin{equation*}
e_{ij}=\overset{k}{\underset{t=1}{\sum }}\left( \overset{k}{\underset{q=1}{%
\sum }}B_{qt}d_{n+k-i-q+1}\right) m_{tj}
\end{equation*}%
and 
\begin{equation*}
e_{(i+r)j}=\overset{k}{\underset{t=1}{\sum }}\left( \overset{k}{\underset{q=1%
}{\sum }}B_{qt}d_{n+k-i-r-q+1}\right) m_{tj}.
\end{equation*}

Using relation $\left( 3.7\right) ,$ it results%
\begin{equation*}
\underset{i\rightarrow \infty }{\lim }\frac{e_{ij}}{e_{(i+r)j}}=\underset{%
i\rightarrow \infty }{\lim }\frac{\overset{k}{\underset{t=1}{\sum }}\left( 
\overset{k}{\underset{q=1}{\sum }}B_{qt}d_{n+k-i-q+1}\right) m_{tj}}{\overset%
{k}{\underset{t=1}{\sum }}\left( \overset{k}{\underset{q=1}{\sum }}%
B_{qt}d_{n+k-i-r-q+1}\right) m_{tj}}=
\end{equation*}%
\begin{equation*}
=\underset{i\rightarrow \infty }{\lim }\frac{\overset{k}{\underset{t=1}{\sum 
}}\left( \overset{k}{\underset{q=1}{\sum }}B_{qt}\left( \overset{k}{\underset%
{s=1}{\sum }}A_{s}\alpha _{s}^{n+k-i-q+1}\right) \right) m_{tj}}{\overset{k}{%
\underset{t=1}{\sum }}\left( \overset{k}{\underset{q=1}{\sum }}B_{qt}\left( 
\overset{k}{\underset{s=1}{\sum }}A_{s}\alpha _{s}^{n+k-i-r-q+1}\right)
\right) m_{tj}}=
\end{equation*}%
\begin{equation*}
=\underset{i\rightarrow \infty }{\lim }\frac{\overset{k}{\underset{q,s=1}{%
\sum }}\Gamma _{sq}\alpha _{s}^{n+k-i-q+1}}{\overset{k}{\underset{q,s=1}{%
\sum }}\Gamma _{sq}\alpha _{s}^{n+k-i-r-q+1}}=
\end{equation*}%
\begin{equation*}
\text{=}\underset{i\rightarrow \infty }{\lim }\frac{\Gamma _{11}\alpha
_{1}^{n+k-i}\text{+}\Gamma _{12}\alpha _{1}^{n+k-i-1}\text{+...+}\Gamma
_{1k}\alpha _{1}^{n-i+1}\text{+}\Gamma _{21}\alpha _{2}^{n+k-i}\text{+}%
\Gamma _{22}\alpha _{2}^{n+k-i-1}\text{+...}}{\Gamma _{11}\alpha _{1}^{n%
\text{+}k\text{-}i\text{-}r}\text{+}\Gamma _{12}\alpha _{1}^{n\text{+}k\text{%
-}i\text{-}r\text{-}1}\text{+...+}\Gamma _{1k}\alpha _{1}^{n\text{-}i\text{-}%
r\text{+}1}\text{+}\Gamma _{21}\alpha _{2}^{n\text{+}k\text{-}i\text{-}r}%
\text{+}\Gamma _{22}\alpha _{2}^{n\text{+}k\text{-}i\text{-}r\text{-}1}\text{%
+...}}\text{=}L.
\end{equation*}%
where $\Gamma _{sq}=A_{s}\left( \overset{k}{\underset{t=1}{\sum }}%
B_{qt}m_{tj}\right) .$ We remark that $\Gamma _{sq}\neq 0,~$due to the
choice of the elements $A_{s},B_{qt}$ and $m_{tj}$. Since $\alpha _{1}>1$
and $\alpha _{1}>\left\vert \alpha _{i}\right\vert ,i\in \{2,...,k\},$ we
obtain that $L=\alpha _{1}^{r}.~$From here, if we consider $i+r\leq k,$ it
results $\frac{e_{ij}}{e_{(i+r)j}}\thickapprox \alpha _{1}^{r},r\in
\{1,2,...,k-1\}._{{}}\medskip $

\textbf{Remark 3.5.} From relation $\left( 3.3\right) ,~$we have $\det
M\cdot \det D_{k}^{n}=\det E,$ then 
\begin{equation}
\left( -1\right) ^{\left( k+1\right) n}a_{k}^{n}\det M=\det E,  \tag{3.7.}
\end{equation}%
by using $\left( 2.13\right) .$

The matrices $D_{k}^{n}$ are used for sending messages. In this way, this
method ensures infinite variants for the chioce of transformation of the
initial matrix $M$.

The method developed above has the property to detect and correct errors in
the submitted message $E$. For this purpose, we used Theorem 3.4 and
relation $\left( 3.7\right) $. From condition $\left( 3.7\right) ,$ we have
the relation between $\det M$ and $\det E$. The determinant of the matrix $M,
$ $\det M,$ is used as a checking element for the code-message matrix $E,$
when we received it using a communication channel. After the matrix $E$ and $%
\det M~$were received, we compute $\det E~$and we check if the relation $%
\left( 3.7\right) $ is satisfied. If the answer is positive, it results that
the matrix $E$ and $\det M~$were transmitted without errors. If the answer
is negative, we have that the matrix $E$ or $\det M$ were received with
errors. For correct the errors, we will use relation $\left( 3.5\right) $.
First, we suppose that we have single error in the received matrix $E$.
Since such an error can appears on the position $\left( i,j\right) ,i,j\in
\{1,2,...,k\}$,$~$it results that we can have $k^{2}$ possibilities. If we
check for all $k^{2}$ possibilities and we don't obtain natural numbers as
solution for the received error, it is possible to have double errors,
triple errors,..., $k^{2}-1$ errors. If in all these situations we do not
obtain positive integer solutions, it results that $\det M$ was sent with
errors or the matrix $E$ has $k^{2}-$fold errors and this matrix is not
correctable. Therefore, the matrix $E$ must be rejected.

This method, in which we use the matrices $D_{k}^{n}$ as encoding matrices
and $D_{k}^{-n}$ as decoding matrices, generalizes the methods developed in
[9] for Fibonacci numbers, in [1] and [2] for Fibonacci $p-$numbers and in
[11] for Fibonacci $k-$numbers to a difference equation of order $k,$ $k\in 
\mathbb{N}$, $k\geq 2,$ given by the relation $\left( 2.9\right) ,$ with $%
a_{1},...a_{k}$ positive integers$.$

The error correcting codes are used widely in modern communications
networks. There are many parameters associated to error correcting codes
which determines the ability of a code to detect and correct errors (Hamming
distance, the rate of a code, etc.) One of these parameters is \textit{the
potential error correction coefficient} $S$, which is the ratio between all
correctable errors and all detectable errors.

We can remark that the code message matrix $E$ can contain single,
double,....,\newline
$k^{2}-$fold errors. Therefore, we have 
\begin{equation*}
\complement _{k^{2}}^{1}+\complement _{k^{2}}^{2}+...+\complement
_{k^{2}}^{k^{2}}=2^{k^{2}}-1
\end{equation*}%
possible errors. Since $k^{2}-$fold errors from the code-message matrix $E$
are not correctable, we can correct $2^{k^{2}}-2$ errors. Therefore we get 
\begin{equation*}
S=\frac{2^{k^{2}}-2}{2^{k^{2}}-1}\thickapprox 1,
\end{equation*}%
that means the correctable possibility of the method is about $100\%.$

This remark generalized the results obtained in [1] and [2] for particular
case of Fibonacci $p-$numbers to a difference equation of degree $k\,$,
defined by $\left( 2.9\right) $.

\begin{equation*}
\end{equation*}

\textbf{Conclusions.} In this paper, we presented some applications of a
difference equation of degree $k$ in Cryptography and Coding Theory.

The algorithm for encryption/decryption messages has some advantages:

i) a block of length $s$ is transformed into a block of different length $r;$

ii) each natural number $n$ has a unique representation using the terms of
this sequence, that means a number cannot have the same encrypted value as
another number;

iii) this method give us a high versatility, ensuring infinite variants for
the chioce of the encrypted keys.

Moreover, using the matrix associated to a difference equation of degree $k$%
\ we have a lot of possibilities to chose a matrix for sending messages,
defining error correcting codes with a very good potential error correction
coefficient.

The above results encourages us to study these equations for finding other
interesting applications of them.

\begin{equation*}
\end{equation*}
\begin{equation*}
\end{equation*}

\textbf{Appendix}%
\begin{equation*}
\end{equation*}

In the following, we \ present some MAPLE procedures used in the encryption
and decryption processes.
\begin{verbatim}
crypt:=proc(st)
local ll,nn,ss,ii,num,n;
num := table(['A'=0, 'B'=1, 'C'=2, 'D'=3, 'E'=4, 'F'=5,
'G'=6, 'H'=7,'I'=8, 'J'=9, K'=10, 'L'=11, 'M'=12, 'N'=13,
'O'=14,'P'=15, 'Q'=16, 'R'=17, 'S'=18, 'T'=19, 'U'=20,
'V'=21, 'W'=22, 'X'=23, 'Y'=24, 'Z'=25, 'x'=26]):
ll := length(st): nn := 1: for ii from 1 to ll do
ss := num[substring(st, ii .. ii)]:
nn := 100* nn+ss: od:
n:=nn-10^(2* ll):print(n):end:
save crypt, `crypt.m`; 
crypt(JOHN)
9140713
\end{verbatim}

\bigskip
\begin{verbatim}
lineq := proc (k, a1, a2, a3, a4, n)
local AA, i, jj, cc, nn, NN, s,j; 
j:=100; AA := array(0 .. j); AA[0] := 0; AA[1] := 0; AA[2] := 0; AA[3] := 1; 
for i from 0 to j-4 do AA[i+4] := AA[i+3]*a1+AA[i+2]*a2+AA[i+1]*a3+AA[i]*a4; 
if n < AA[i+4] and AA[i+3] < n then jj := i+3 end if;
s := jj-k+2 end do; cc := array(1 .. s); cc[1] := trunc(n/AA[jj]); 
nn := n-cc[1]*AA[jj]; for i from 2 to s do cc[i] := trunc(nn/AA[jj-i+1]);
nn := nn-cc[i]*AA[jj-i+1] end do; NN := 10^(2*s-2)*cc[1]; 
for i from 2 to s do NN := NN+10^(2*s-2*i)*cc[i] end do;print(s); print(NN);
end proc; save lineq, `lineq.m`; lineq(4, 18, 10, 13, 3, 9140713):
6
40418081505
\end{verbatim}

\bigskip
\begin{verbatim}
decrypt:=proc(nn) 
local alpha, ss,mm,rr,ii, ans,A,B,C,D,E,F,G,H,II,J,K,L,M,
O,P,Q,R,S,T,U,V,W,X,Y,Z,x;
alpha:= table([0 = A, 1 = B, 2 = C, 3 = D, 4 = E, 
5 = F, 6 = G, 7 = H, 8 = I, 9 = J, 10 = K, 11 = L,
12 = M, 13 = N, 14 = O,15=P,16=Q,17=R,18=S,19=T,
20=U,21=V,22=W,23=X,24=Y,25=Z,26=x]):
mm := nn: rr:=floor(trunc(evalf(log10(mm)))/2)+1: ans:=` `:   
for ii from 1 to rr do mm:=mm/100: ss:=alpha[frac(mm)*100]:
ans:=cat(ss,ans): mm:=trunc(mm) od: ans;  
end: save decrypt, `decrypt.m`; 
decrypt(40418081505);
EESIPF
\end{verbatim}

\bigskip
\begin{verbatim}
decryptM:=proc(nn) 
local alpha, ss,mm,rr,ii, ans,A,B,C,D,E,F,G,H,II,J,K,L,M,
O,P,Q,R,S,T,U,V,W,X,Y,Z,x;
alpha:= table([0 = A, 1 = B, 2 = C, 3 = D, 4 = E, 
5 = F, 6 = G, 7 = H, 8 = I, 9 = J, 10 = K, 11 = L,
12 = M, 13 = N, 14 = O,15=P,16=Q,17=R,18=S,19=T,
20=U,21=V,22=W,23=X,24=Y,25=Z,26=x,32=y,121=z]):
mm := nn: rr:=floor(trunc(evalf(log10(mm)))/3)+1: ans:=` `:   
for ii from 1 to rr do mm:=mm/1000: ss:=alpha[frac(mm)*1000]:
ans:=cat(ss,ans): mm:=trunc(mm) od: ans;  
end: save decryptM, `decryptM.m`; 
decryptM(23032121012001003);
XyzMBD 
\end{verbatim}

\bigskip
\begin{verbatim}
delineq := proc (st, a1, a2, a3, a4)
local AA, BB, cc, ll,nn, num, s,ii; 
num := table(['A'=0, 'B'=1, 'C'=2, 'D'=3, 'E'=4, 'F'=5,
'G'=6, 'H'=7,'I'=8, 'J'=9, K'=10, 'L'=11, 'M'=12, 'N'=13,
'O'=14,'P'=15, 'Q'=16, 'R'=17, 'S'=18, 'T'=19, 'U'=20,
'V'=21, 'W'=22, 'X'=23, 'Y'=24, 'Z'=25, 'x'=26]):
ll := length(st):  BB:=array(1..ll):AA:=array(0..ll+2):
for ii from 1 to ll do BB[ii]:= num[substring(st, ii .. ii)]  od:
AA[0] := 0: AA[1] := 0: AA[2] := 0: AA[3] := 1:
for ii from 0 to ll-2  do 
AA[ii+4] := AA[ii+3]*a1+AA[ii+2]*a2+AA[ii+1]*a3+AA[ii]*a4;od:
nn[]:=0: for ii from 1 to ll do nn:=nn+BB[ii]*AA[ll+3- ii]: od:
print(nn);end proc; save delineq, `delineq.m`;
delineq(EESIPF,18, 10, 13, 3): 
9140713                                                              
\end{verbatim}

\bigskip
\begin{verbatim}
split := proc (st, k)
local ll, ii, rr, AA; 
ll := length(st); rr := 0; AA := array(1 .. ll/k); 
for ii to ll/k do AA[ii] := substring(st, ii+rr .. ii+k-1+rr); 
rr := rr+k-1 end do; print(AA); 
end proc; save split, `split.m`; 
split(EESIPFMDENBNMCMHNNGKKBDG, 6);
[EESIPF MDENBN MCMHNN GKKBDG]                                   
\end{verbatim}

\bigskip

\begin{equation*}
\end{equation*}

\textbf{Acknowledgments.} The author thanks the referees for their
suggestions and remarks which helped me to improve this paper.%
\begin{equation*}
\end{equation*}

\textbf{References}%
\begin{equation*}
\end{equation*}

[1] Basu, M., Prasad, B., \textit{The generalized relations among the code
elements for Fibonacci coding theory}, Chaos, Solitons and Fractals,
41(2009), 2517-2525.

[2] Basu, M., Prasad, B., \textit{Coding theory on the m-extension of the
Fibonacci p-numbers,} Chaos, Solitons and Fractals, 42(2009), 2522--2530.

[3] Brown, Jr.J. L., \textit{Note on Complete Sequences of Integers}, The
American Mathematical Monthly, (68)(6)(1961), 557-560.

[4] Esmaeili, M., Esmaeili, Mo., \textit{A Fibonacci-polynomial based coding
method with error detection and correction}, Computers and Mathematics with
Applications, 60(2010), 2738-2752.

[5] Sergio Falcon, S., Plaza, A., \textit{On k-Fibonacci numbers of
arithmetic indexes}, Applied Mathematics and Computation, 208(2009),
180--185.

[6] Hoggatt, Jr.V. E., King, C., \textit{Problem E1424}, Amer. Math.
Monthly, Vol. 67(1960), p. 593.

[7] Johnson, R.C., \textit{Fibonacci numbers and matrices},\newline
http://maths.dur.ac.uk/\symbol{126}dma0rcj/PED/fib.pdf

[8] Klein, S.T., Ben-Nissan, M.K., \textit{On the Usefulness of Fibonacci
Compression Codes, The Computer Journal}, 2005, doi:10.1093/comjnl/bxh000

[9] Stakhov, A.P., \textit{Fibonacci matrices, a generalization of the
\textquotedblleft Cassini formula\textquotedblright , and a new coding theory%
}, Chaos, Solitons and Fractals, 30(2006), 56-66.

[10] Stakhov, A.P., \textit{The \textquotedblleft golden\textquotedblright\
matrices and a new kind of cryptography}, Chaos, Solitons and Fractals,
32(2007), 1138--1146.

[11] Tahghighi, M., Jafaar, A., Mahmod, R., Said, M. R., \textit{%
Generalization of Golden Cryptography based on }$k$\textit{-Fibonacci Numbers%
}, International Conference on Intelligent Network and Computing (ICINC
2010), 2010.

[12] Yeates, A. C., \textit{Application of Linear Sequences to Cryptography}%
, Honors Theses, Paper 191, 2013.

[13] Zeckendorf, E., \textit{Repr\'{c}sentation des nombres naturels par une
somme de nombres de Fibonacci ou de nombres de Lucas}, Bull. Soc. R. Sci. Li%
\'{c}ge, 41(1972), 79-182.%
\begin{equation*}
\end{equation*}

\begin{center}
\bigskip
\end{center}

\end{document}